\documentclass[sn-aps]{sn-jnl}
\usepackage{multirow}
\usepackage[normalem]{ulem} 
\usepackage{enumerate}
\usepackage{footmisc}

\usepackage{graphicx}%
\usepackage{multirow}%
\usepackage{amsmath,amssymb,amsfonts,amsbsy}
\usepackage{amsthm}%
\usepackage{mathrsfs}%
\usepackage[title]{appendix}%
\usepackage{xcolor}%
\usepackage{textcomp}%
\usepackage{manyfoot}%
\usepackage{booktabs}%
\usepackage{algorithm}%
\usepackage{algorithmicx}%
\usepackage{algpseudocode}%
\usepackage{listings}%
\begin{document}
\title[Mesonic modes in confining model at finite temperature]{Mesonic modes in confining model at finite temperature}


\author[1]{\fnm{A.E.} \sur{Radzhabov}}\email{aradzh@icc.ru}
\equalcont{These authors contributed equally to this work.}

\author*[2]{\fnm{X.L.} \sur{Shang}}\email{shangxinle@impcas.ac.cn}

\affil[1]{ \orgname{Matrosov Institute for System Dynamics and Control Theory SB RAS}, \orgaddress{\city{Irkutsk}, \postcode{664033}, \state{State}, \country{Russia}}}

\affil*[2]{\orgname{Institute of Modern Physics Chinese Academy of Sciences}, \orgaddress{ \city{Lanzhou}, \postcode{730000}, \country{China}}}


\abstract{
The mass spectrum of pseudoscalar and scalar meson modes at finite temperature is studied in the framework of a nonlocal quark model.
The model implements quark confinement via the modification of the Laplace transform of the quark propagator.
In order to synchronize the confining and deconfining phases, a modification of the transform is proposed.
The behavior of the screening masses of mesons is studied in a wide region of temperatures, while the pole masses are described up to the deconfining phase transition.
}

\maketitle

\section{Introduction}
One of the most interesting features of strong interactions is confinement, which prevents the appearance of fundamental color degrees of freedom in the physically observed mass spectrum.
The only {\it ab initio} method describing confinement is lattice QCD calculations.
As lattice QCD calculations produce numerical data, effective modeling provides a complementary framework for theoretical understanding of dynamics \cite{Brambilla:2014jmp,Gross:2022hyw}.

The symmetries of strong interactions provide the basis for constructing effective models such as the Nambu–Jona-Lasinio (NJL) model \cite{Nambu:1961tp} and Chiral Perturbation Theory (ChPT) \cite{Gasser:1983yg}.
One of the important symmetries of strong interactions is chiral symmetry, which is spontaneously broken by the quark condensate.
The NJL model incorporates the feature of spontaneous and explicit chiral symmetry breaking by the quark condensate and the current quark masses, as in QCD.
On the other hand, the NJL model operates with a constant constituent quark mass, which seems a rather rough approximation.

In the nonlocal version of the NJL model \cite{GomezDumm:2001fz, Dorokhov:2000gu}, the quark mass function is momentum-dependent, which makes it similar to Dyson–Schwinger studies of QCD with full quark and gluon (ghost) propagators and their interactions  \cite{Roberts:2000aa}. 
In order to make calculations tractable, an approximation is required, typically the rainbow-ladder scheme for the Dyson–Schwinger equation, together with the corresponding Bethe–Salpeter equation for mesonic bound states.
The feature of quark confinement can be implemented in the nonlocal NJL model by requiring that the quark propagator (or a part of it) be an entire function \cite{Radzhabov:2003hy}. A similar scheme is also used in DSE studies \cite{Burden:1995ve, Hecht:2000xa,Chen:2021guo}.
However, the extension of this scheme to finite temperature is not a simple task since at some point the phase transition to quark matter should occur and quark should appear as a real degrees of freedom.
This type of scheme is sometimes called technical or analytical confinement, since it does not follow from first principles \cite{Efimov:1993zg}.

In our previous work, a modification was suggested that involves a cutoff of a part of the quark propagator after the inverse Laplace transform \cite{Radzhabov:2024cvo}.
On the other hand, to model the deconfinement phase transition at a certain temperature, this cutoff should be removed for a real quark pole.
Due to this, there was a first-order phase transition at the phase transition point.
In order to remove this unwanted feature, it is suggested that the four-quark coupling constants be different in the hadronic phase and quark matter, so that the phase transition at finite temperature becomes second order.
Here, we consider another possibility: how one can use the modification of the Laplace transform to synchronize different phases.
In addition, we discuss the mass spectra of pions and sigma mesons at finite temperatures.

\section{Model}
The $SU(2)$ nonlocal chiral quark model with separable interaction in the pseudoscalar–scalar channels is described by the Lagrangian:
\begin{align}
&\mathcal{L}
 = \bar{q}(x)(i \hat{\partial}-M_c)q(x)+
\frac{G}{2}\bigg(\Big(J_S^a(x)\Big)^2+\Big(J^a_P(x)\Big)^2\bigg) ,\quad %
\label{Lagr} 
\end{align}
$M_c = \mathrm{diag}(m_c, m_c)$ is the current quark mass matrix, $G$ is the four-quark coupling constant, and $J_{S,P}^a(x)$ denote the quark currents. %
Following the NJL model \cite{Nambu:1961tp}, the interaction kernel is taken to be chirally symmetric. However, its current–current interaction is taken in a nonlocal one-gluon-exchange-like (OGE) form \cite{Ito:1991sz,Schmidt:1994di,GomezDumm:2006vz,Hell:2008cc}:
\begin{align}
J_{M}^{a}(x) 
= \int d^4x_1 d^4x_2 \,
\delta\left(x-\frac{x_1+x_2}{2}\right)
g((x_1-x_2)^2)\,
\bar{q}(x_1) \, \Gamma_{M}^{a} \, q(x_2), \label{eq2} %
\end{align}
with $M=S,P$ and  %
$\Gamma_{{S}}^{a}=\lambda^{a}$, $\Gamma_{{P}}^{a}=i\gamma^{5}\lambda^{a}$. %
For the $SU(2)$ model,  
the flavour matrices are Pauli matrices: $\lambda^{a}\equiv\tau^{a}$, $a=0,...,3$ with $\tau^0=1$.
%
%
The nonlocal form factor is taken in the Gaussian form, $g(k^2)=\exp(-k^2/\Lambda^2)$ in momentum space. The parameter $\Lambda$ is related to the characteristic scale of spontaneous symmetry breaking\footnote{The numerical calculations are performed using the parameter set from Scheme II of \cite{GomezDumm:2006vz}, which yields $\langle \bar{q}q \rangle = -(240~\text{MeV})^3$. The corresponding values are $m_c = 5.8$ MeV, $m_d = 424$ MeV, and $\Lambda = 752.2$ MeV.}.

In the mean-field approximation, the thermodynamic potential of the nonlocal model in a medium is \cite{Radzhabov:2010dd,Carlomagno:2019yvi}
\begin{align}
\Omega^{MF} =\frac{m_d^2}{2 G} -4 \sum_{i=0,\pm}\int_{k,n}\ln\bigg[k_{n,i}^2+m^2(k_{n,i}^2)\bigg] 
+ U (\Phi,\bar{\Phi}) + \Omega_0, \label{ThermodynamicPoten}
\end{align}
where $m(k^2)=m_c+m_d g(k^2)$, $k_{n,i}=(\omega_n^i-i {\mu})^2+\mathbf{k}^2$ and $\int_{k,n} \equiv T\sum_n\frac{d^3k}{(2\pi)^3}$. Fermionic Matsubara frequencies $\omega_n^i$ are partially shifted due to the presence of a Polyakov loop: $\omega_n^{\pm}=\omega_n\pm \phi_3$, $\omega_n^0=\omega_n$ and $\omega_n=(2n+1)\pi T$.
The thermodynamic potential is divergent due to the current quark contribution.
The infinite normalization constant $\Omega_0$ is chosen such that the pressure vanishes under vacuum conditions, i.e., at zero temperature, zero chemical potential, and the vacuum value of $m_d$.
There are different versions of the Polyakov loop potential $U(\Phi, \bar{\Phi})$. In this paper, the modified logarithmic Polyakov loop potential is used \cite{Contrera:2012wj}.

In the presence of the Polyakov loop, the equations of motion for the mean fields are:
\begin{align}
\frac{\partial \Omega^{MF}}{\partial m_d} = 0 , \quad \frac{\partial \Omega^{MF}}{\partial \phi_3} = 0.
\label{GapEqAtT}
\end{align}
The first of these equations \eqref{GapEqAtT}, i.e., the gap equation for $m_d$, has the form:
\begin{align}
\frac{\partial \Omega^{MF}}{\partial m_d} =\frac{m_{d}}{G}-8 \sum_{i=0,\pm}\int_{k,n} 
g(k_{n,i}^2)m(k_{n,i}^2)\mathrm{D}(k_{n,i}^2)\label{GapT},
\end{align}
where $\mathrm{D}(k_{n,i}^2)=(k_{n,i}^2+m^2(k_{n,i}^2))^{-1}$  is the scalar propagator.

\section{Confining prescription}
The analytical structure of the scalar propagator is crucially dependent on the particular form of $g(k^2)$. For the Gaussian case, it has infinitely many singularities, which is similar to DSE studies \cite{,Dorkin:2014lxa,Windisch:2016iud}.
The scalar propagator for the case with only pole singularities can be represented in the form:
\begin{align}
&\mathrm{D}(k^2)= \sum\limits_{i=1}^{\infty} \frac{R_{z_i} }{k^2+z_i} \label{SeriesMomentum1} ,
\end{align} 
where $z_i$ are the pole positions with corresponding residues $R_{z_i}$. For convenience, the poles are ordered starting from the smallest in absolute value.

The general idea for modeling quark confinement is to modify this part of the quark propagator in order to remove singularities. In our previous paper \cite{Radzhabov:2025pyx}, this step is performed using the prescription of the inverse Laplace transform $\mathbf{D}(\alpha)$ of the momentum-space propagator $\mathrm{D}(k^2)$. The cut of the transform
\begin{align}
\mathbf{D}_R(\alpha)=\mathbf{D}(\alpha)\theta\bigl({1}/{\Lambda_c^2}-\alpha\bigr)%
\end{align}
leads to an entire function $\mathrm{D}_R(k^2)$ in momentum space:
\begin{align}
\mathrm{D}_R(k^2)=
\mathrm{D}(k^2) - \sum\limits_{i=1}^{\infty} \frac{R_{z_i} 
Q\left(k^2+z_i\right)
 }{k^2+z_i}.\label{InvDRCon} 
\end{align} 
The scale parameter $\Lambda_c$ can be associated with the confinement scale.
In order to realize deconfinement, one can start the subtraction sum in \eqref{InvDRCon} from two, since the pole for $i=1$ corresponds to the physical quark pole, and $\mathrm{D}_R(k^2)$ can be rewritten in the form
\begin{align}
\mathrm{D}_R(k^2)=
%
%
\frac{R_{z_1}}{k^2+z_1-i\epsilon}+\bigg[%
\mathrm{D}(k^2)-\frac{R_{z_1}}{k^2+z_1}
-\sum\limits_{i=2}^{\infty} \frac{R_{z_i} 
Q\left(k^2+z_i\right)
}{k^2+z_i}\bigg],\label{InvDRdecon} 
\end{align} 
where the bracketed expression is an entire function.

\begin{figure}[ht]
    \centering
        \includegraphics[width=0.45\textwidth]{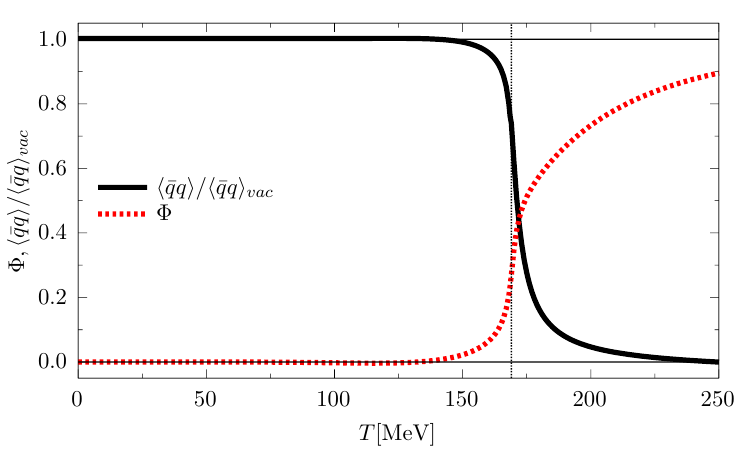}
\caption{Quark condensate normalized to its vacuum value (solid black line) and Polyakov loop (red dashed line) as functions of temperature. The thin vertical dashed line denotes the critical temperature. }
\label{FiniteTCondPolLoop}
\end{figure}
\begin{figure}[ht]
    \centering
        \includegraphics[width=0.45\textwidth]{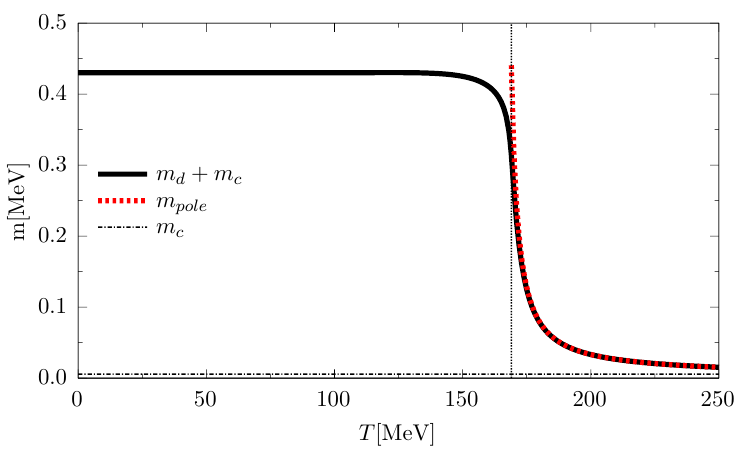}
\caption{Finite-temperature behavior of the current quark mass, $m_d+m_c$ (where $m_d$ is the mass scale), and the quark pole mass. The thin vertical dashed line denotes the critical temperature.}
\label{FiniteTMd}
\end{figure}

With such a simple prescription, a problem appears because the gap equations for the confining and deconfining phases are not synchronized, and there is a jump between the phases.
In our previous work \cite{Radzhabov:2025pyx}, it was suggested to use two different values of the four-quark coupling constant in order to compensate for the difference between the integrands in the gap equations \eqref{GapT}.
In the present paper, a different idea is suggested — namely, one can modify the Laplace transform of the scalar propagator in order to make the gap equation smooth with respect to $m_d$. The guiding principle is related to the normalization of the mean-field gap equation of the original model without the confining extension. The details are given in the appendix.
Additionally, as an approximation, the value of $\phi_3$ is taken to be the same as in the original model.
The quark condensate and Polyakov loop are given in Fig.~\ref{FiniteTCondPolLoop}.
Fig.~\ref{FiniteTMd} shows the behavior of the sum $m_d + m_c$. Above the critical temperature, a pole appears in the quark propagator, and the corresponding behavior of the quark mass is also shown in Fig.~\ref{FiniteTMd}.

\section{Meson loops at finite T}

The pion polarization loop at finite $T$ and $\mu$ has the form\footnote{For the pion properties, a medium-dependent $Z$ factor is introduced to reproduce the pion screening mass at a given $T$, as in the model without the confinement prescription. As in \cite{Radzhabov:2025pyx}, where the $Z$ factor was introduced for the pion in vacuum, it can be made slightly $T$-dependent. Details are given in appendix.}:
\begin{align}
&\Pi_{\pi}((\nu_M-i\mu_{ab})^2,\mathbf{p}^2)={8 } \sum_{i=0,\pm}\int_{k,n}  g^2(q^2)\, \mathrm{D}(k_-^2)\mathrm{D}(k_+^2) \times \notag \\
&\times \left[ 
{(k_- \cdot k_+)+m(k_+^2)m(k_-^2)} \right]
,
\label{poloperOGE} %
\end{align}
where quark momenta are %
\begin{align}
k_+&=(\omega_n^i-i {\mu}_a+\nu_M,\mathbf{k}+\zeta \mathbf{p}) \notag \\
k_-&=(\omega_n^i-i {\mu}_b,\mathbf{k}-(1-\zeta) \mathbf{p})
\end{align}
where $a$ and $b$ are quark flavors, and $\mu_{ab}= \mu_a - \mu_b$ is the meson chemical potential. 
The relations $p=k_+-k_-$ and $q=(k_++k_-)/2$ for the four-momenta hold. 
In the general case, the flow of three-momentum in the diagram is arbitrary, with $0<\zeta<1$. For a scalar meson, the plus sign in brackets should be changed to a minus.

\begin{figure}[t]
    \centering
        \includegraphics[width=0.45\textwidth]{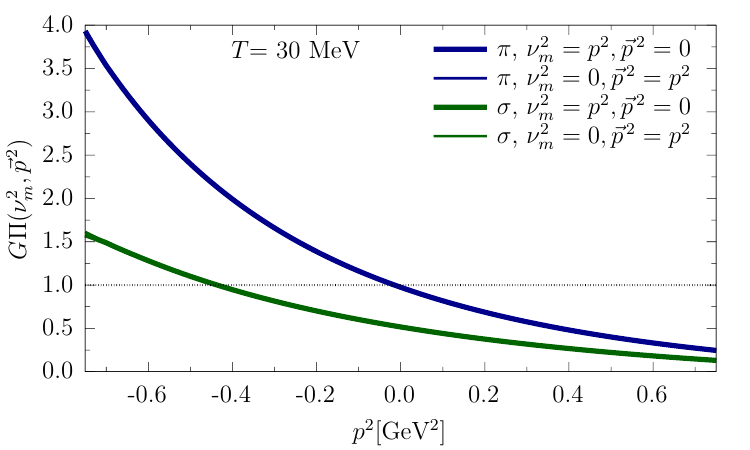}\\
        \includegraphics[width=0.45\textwidth]{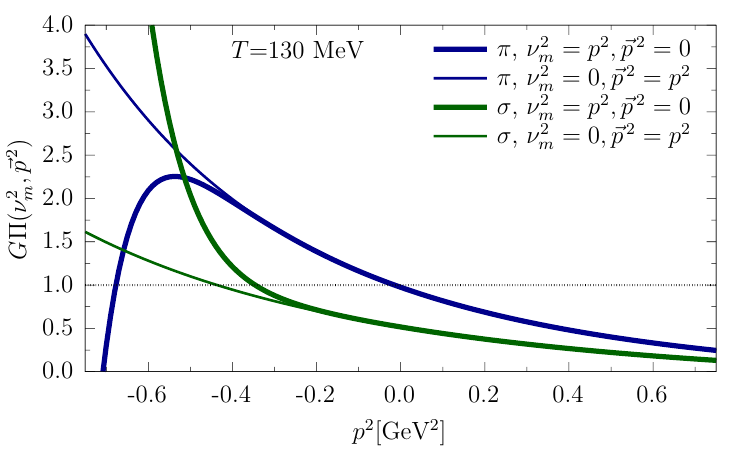}\\
        \includegraphics[width=0.45\textwidth]{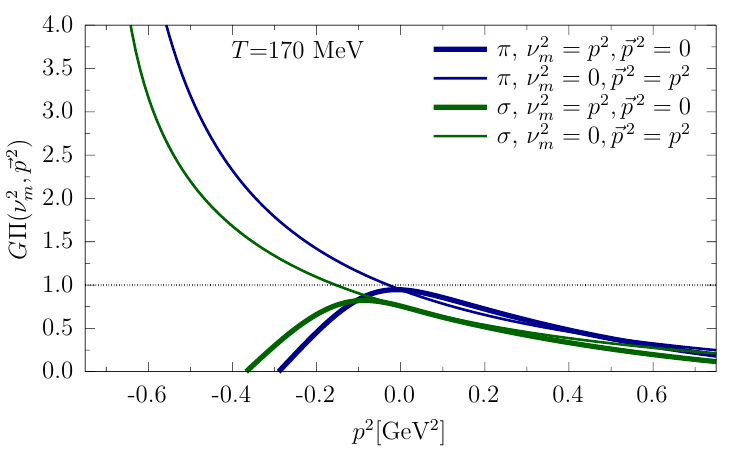}
\caption{Pion (blue) and sigma-meson (green) polarization loops multiplied by the four-quark coupling constant at finite temperatures $T=30$, $130$, and $170$ MeV. Thick (thin) lines correspond to temporal (spatial) correlations. The thin dashed horizontal line denotes $1$, which corresponds to the meson pole.
}
\label{PolLoopFinT}
\end{figure}

At finite temperature, Lorentz invariance is broken due to the appearance of the heat bath collective motion vector $u$. 
Therefore, the dependence of correlations on the directions parallel and perpendicular to this vector should be studied. It is convenient to choose this vector as $u=(1,\mathbf{0})$, and therefore consider the zero component and three-momentum separately. One can consider two limiting cases: 
\begin{itemize}
\item In the first case, the meson has only the zero component of momentum (or the fourth component in the Euclidean formulation).
It corresponds to an analytic continuation from real Matsubara frequencies to complex values $\nu_M=iM_\pi^{\mathrm{pol}}$:
\begin{align}
-G^{-1}+\Pi_{\pi}(-(M_\pi^{\mathrm{pol}})^2,0)=0. \label{pionMassPole}
\end{align}
This case is referred to as the pole, temporal, or dynamical mass \cite{Hufner:1994ma,Florkowski:1993br,Florkowski:1997pi,Ishii:2016dln}.
\item 
The other case, that of screening masses of mesons, is connected with the zero energy component, i.e., the zero Matsubara mode, and the polarization loop is analytically continued to complex values of the three-momentum squared $\mathbf{p}^2=-(M_\pi^{\mathrm{scr}})^2$:
\begin{align}
-G^{-1}+\Pi_{\pi}(0,-(M_\pi^{\mathrm{scr}})^2)=0. \label{pionMassScreen}
\end{align}
This definition is connected with the asymptotic behavior of the correlation function and corresponds to the screening of the potential by meson exchange in coordinate space, $\exp(-M_\pi^{\mathrm{scr}} r)$,  \cite{Blaschke:2000gd,Contrera:2009hk,Benic:2013eqa,Ishii:2013kaa,Ishii:2016dln,Carlomagno:2019yvi,Dumm:2021vop,Chen:2024emt}.
\end{itemize}
In the limit of zero temperature, the masses should coincide: $M_\pi^{\mathrm{pol}}=M_\pi^{\mathrm{scr}}$.
In lattice calculations, usually only screening masses are calculated. The pole masses are sensitive to the lattice size $N_t$, and it is hard to estimate them reliably.
The behavior of the meson polarization loops is shown in Fig.~\ref{PolLoopFinT} for $T=30$, $130$, and $170$ MeV. Two cases are considered: longitudinal and transverse with respect to the heat bath motion, i.e., the temporal and screening correlation functions. At low temperature ($T=30$ MeV), Lorentz-breaking effects are almost invisible, and the corresponding masses coincide. As the temperature increases, the behavior of the temporal polarization loops for large time-like momentum starts to change, and as a result, the pole mass of the scalar meson is modified. In Fig.~\ref{PolLoopFinT} for $T=130$ MeV, one can see this situation, and, additionally, a second heavy solution appears in the pseudoscalar channel. However, since the considered model includes only the ground states of mesons, the appearance of a second solution cannot be discussed on such simple grounds. After deconfinement ($T=170$ MeV), the temporal correlation functions change their behavior drastically, and no real-valued solution can be found. The complex-valued solution, i.e., unstable mesonic states after deconfinement, will be studied in a separate paper.

\section{Lattice}

\begin{figure}[ht]
    \centering
        \includegraphics[width=0.45\textwidth]{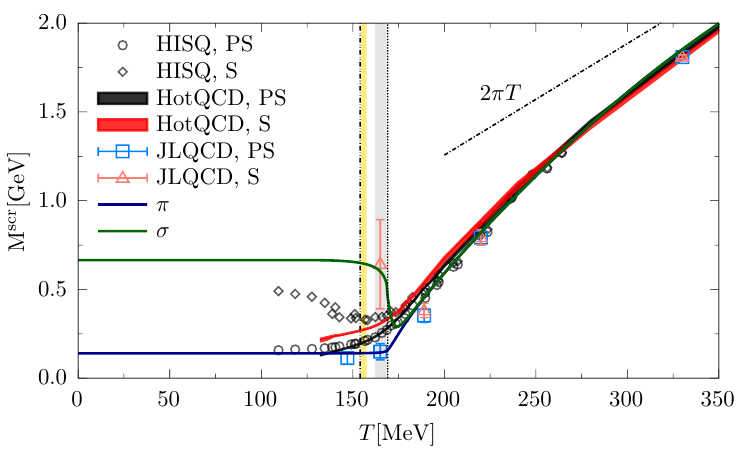}
\caption{Behavior of screening meson masses as a function of the temperature: blue lines correspond to the pion and green lines to the $\sigma$-meson in the nonlocal model; shaded curves represent HotQCD calculations for pseudoscalar (black) and scalar (red) mesons; circles (pseudoscalar) and diamonds (scalar) correspond to calculations with the HISQ action, while boxes (pseudoscalar) and triangles (scalar) are the JLQCD results; the dash-dotted line is the asymptotic $2\pi T$.
The thin vertical dashed line denotes the critical temperature in the nonlocal model; the grey shaded region corresponds to the JLQCD result, the yellow shaded region to the HotQCD result, and the dash-dotted line to the central value from calculations with the HISQ action.}
\label{MassesLargeScaleComparisonLattice}
\end{figure}

Screening correlations at finite temperatures have been extensively studied in lattice calculations; see, for example, \cite{Cheng:2010fe,HotQCD:2012vvd,Aoki:2012yj,Cossu:2013uua,Buchoff:2013nra,Maezawa:2016pwo,Murakami:2019yre,Bazavov:2019www,Ding:2020xlj,Aoki:2025mue}.
A comparison of the model calculations of the pion and $\sigma$-meson screening masses with lattice results \cite{Maezawa:2016pwo,Bazavov:2019www,Aoki:2025mue} is shown in Fig.~\ref{MassesLargeScaleComparisonLattice} as a function of temperature.
The continuum extrapolation was performed by the HotQCD collaboration \cite{Bazavov:2019www}, while results obtained with the HISQ action \cite{Maezawa:2016pwo} and by the JLQCD collaboration \cite{Aoki:2025mue} are presented for different lattice sizes. The model reproduces the pion screening mass in reasonable agreement with lattice data. For the scalar meson, due to the restoration of chiral symmetry, the large-$T$ behavior is similar to that of the pion and therefore consistent with lattice results, while the low-$T$ results are found to agree with the JLQCD data, unlike the HotQCD data.
The difference in the scalar channel on the lattice at the lowest simulated points, as discussed in \cite{Aoki:2025mue} for the HotQCD calculations, tends to twice the pion mass in the low-$T$ limit due to a lattice artifact of the staggered quark action \cite{Bazavov:2019www}. In \cite{Aoki:2025mue}, which uses the M\"{o}bius domain-wall fermion formulation, this artifact is absent.

\begin{figure}[ht]
    \centering
        \includegraphics[width=0.45\textwidth]{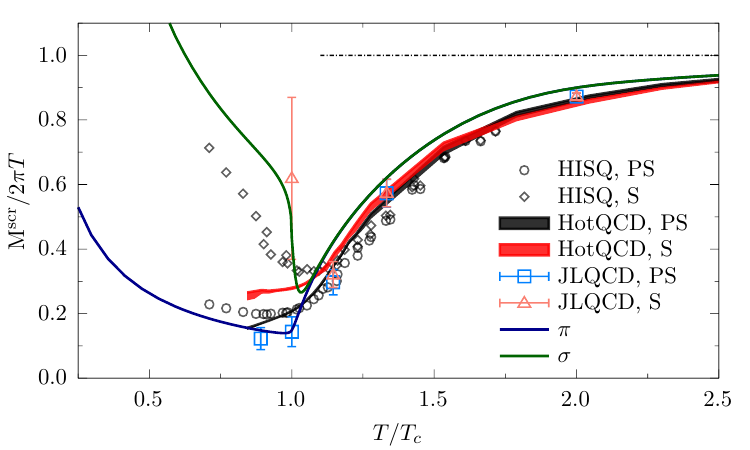}
\caption{Behavior of screening meson masses scaled to the asymptotic limit, $M/2\pi T$, as a function of the scaled temperature $T/T_c$: blue lines correspond to the pion and green lines to the $\sigma$-meson in the nonlocal model; shaded curves represent HotQCD calculations for pseudoscalar (black) and scalar (red) mesons; circles (pseudoscalar) and diamonds (scalar) correspond to calculations with the HISQ action, while boxes (pseudoscalar) and triangles (scalar) are the JLQCD results.}
\label{MassesLargeScaleComparisonLatticeVsTdivTcV2}
\end{figure}

Since the critical temperature is slightly different for these calculations — HotQCD: $156\pm1.5$ MeV, HISQ: $154\pm9$ MeV \cite{Bazavov:2011nk}, JLQCD: $165\pm3$ MeV, and $169$ MeV in the nonlocal model — it is interesting to examine the situation as a function of the scaled temperature $T/T_c$ in Fig.~\ref{MassesLargeScaleComparisonLatticeVsTdivTcV2}. Additionally, the masses are scaled by $2\pi T$, which represents the asymptotic value at high $T$. It can be seen that the difference in critical temperatures does not significantly affect the agreement between the model results and lattice calculations.
 
%

\section{Screening vs pole masses}

In effective models, simultaneous calculations of pole and screening masses are not a trivial task. This is due to the nonperturbative nature of QCD.
There are the following problems: in NJL-like models, the regularization should be Lorentz invariant from the beginning, i.e., one cannot use a three-dimensional cutoff scheme. 
In nonlocal models or DSE studies, the Matsubara summation is problematic due to the complicated structure of the quark propagator. Confinement should also be included in the hadronic phase, with a transition to quark matter at high temperature. 

Therefore, in different models, only the temporal \cite{Hansen:2006ee} or only the screening \cite{Blaschke:2000gd,Contrera:2009hk,Benic:2013eqa,Ishii:2013kaa,Carlomagno:2019yvi,Dumm:2021vop,Chen:2024emt,RamirezGarrido:2025rsu} masses are calculated.
Both the pole and screening masses are estimated in the PV-regularized NJL model \cite{Florkowski:1993br} (pion and $\sigma$), the PNJL model \cite{Ishii:2016dln} (scalar and pseudoscalar mesons), and the quark-meson model \cite{Helmboldt:2014iya} (for the pion).
In the confining model, the calculation of both modes is possible.

\begin{figure}[ht]
    \centering
        \includegraphics[width=0.45\textwidth]{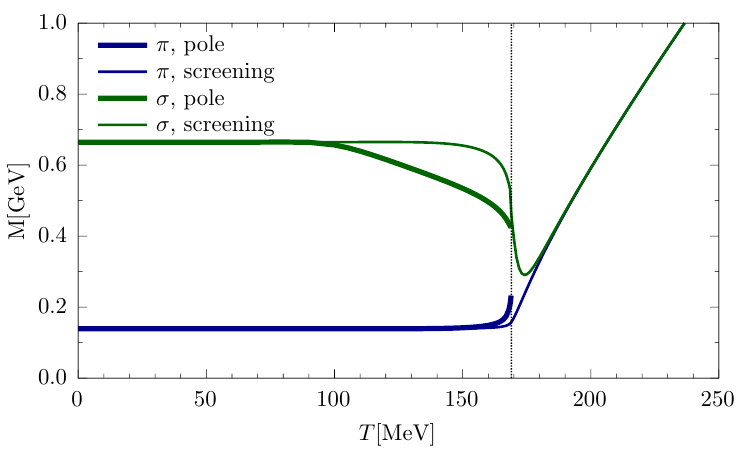}
\caption{Finite-temperature behavior of the pole (thick) and screening (thin) meson masses: blue lines correspond to the pion and green lines to the $\sigma$-meson. The thin vertical dashed line denotes the critical temperature.}
\label{MesonMasses}
\end{figure}

The results of our calculations for the pion and sigma pole and screening masses are shown in Fig.~\ref{MesonMasses}. One can see that at low $T$, the pole and screening masses coincide. Around $100$ MeV, the pole mass of the $\sigma$-meson starts to decrease. Above $\sim 140$ MeV, the pole mass of the $\sigma$-meson continues to decrease, while the pion pole mass begins to increase. The pion screening mass starts to rise near the phase transition. After the phase transition, the screening masses of the pion and sigma become degenerate. However, for the pole masses, no real solution can be found after $T_c$, and we plan to investigate the complex-valued poles in future work.

\begin{figure}[ht]
    \centering
        \includegraphics[width=0.45\textwidth]{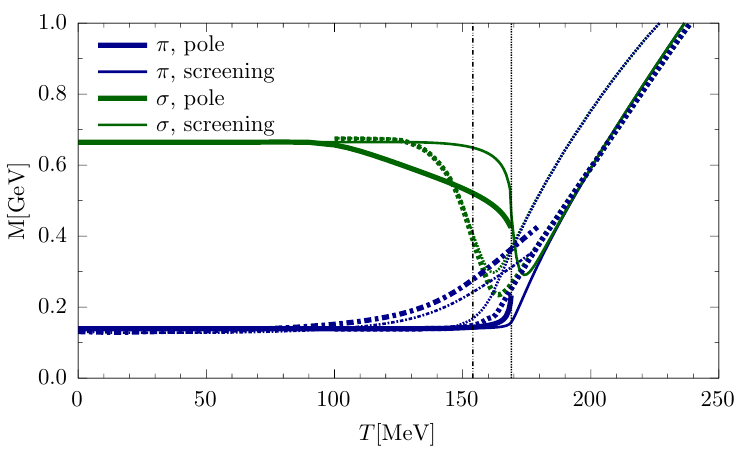}
\caption{Finite-temperature behavior of the pole (thick) and screening (thin) meson masses: blue lines correspond to the pion and green lines to the $\sigma$-meson. Solid lines correspond to the nonlocal model, dashed lines to the NJL calculations with PV regularization \cite{Ishii:2016dln}, and dash-dotted lines to the calculations in the quark-meson model \cite{Helmboldt:2014iya}. The thin vertical dashed line denotes the critical temperature in the nonlocal model, while the dash-dotted line corresponds to the value used in \cite{Ishii:2016dln}.}
\label{MassesComparisonV2}
\end{figure}

\begin{figure}[ht]
    \centering
        \includegraphics[width=0.45\textwidth]{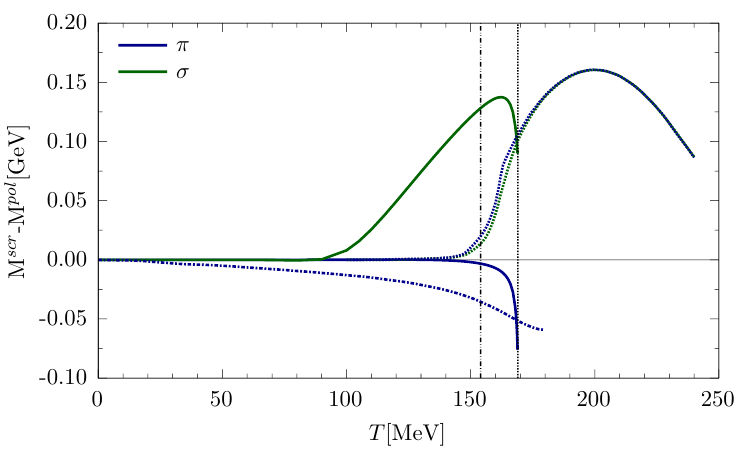}
\caption{Difference between the screening and pole meson masses at finite temperature: blue lines correspond to the pion and green lines to the $\sigma$-meson. Solid lines correspond to the nonlocal model, dashed lines to the NJL calculations with PV regularization \cite{Ishii:2016dln}, and dash-dotted lines to the calculations in the quark-meson model \cite{Helmboldt:2014iya}. The thin vertical dashed line denotes the critical temperature in the nonlocal model, while the dash-dotted line corresponds to the value used in \cite{Ishii:2016dln}.}
\label{MassesDiffComparison}
\end{figure}

\begin{figure}[ht]
    \centering
        \includegraphics[width=0.45\textwidth]{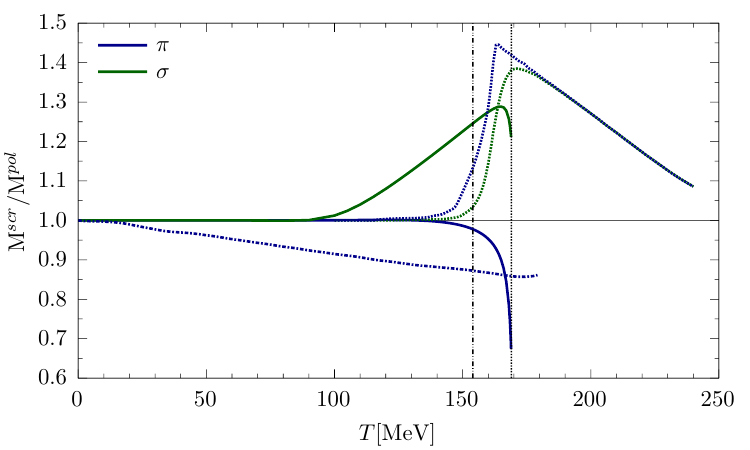}
\caption{Ratio of the screening to pole meson masses at finite temperature: blue lines correspond to the pion and green lines to the $\sigma$-meson. Solid lines correspond to the nonlocal model, dashed lines to the NJL calculations with PV regularization \cite{Ishii:2016dln}, and dash-dotted lines to the calculations in the quark-meson model \cite{Helmboldt:2014iya}. The thin vertical dashed line denotes the critical temperature in the nonlocal model, while the dash-dotted line corresponds to the value used in \cite{Ishii:2016dln}.}
\label{MassesRatioComparison}
\end{figure}

A comparison of calculations in the confining model of the pole and screening masses with those from the PNJL model with PV regularization \cite{Ishii:2016dln} and the quark-meson model \cite{Helmboldt:2014iya} is shown in Fig.~\ref{MassesComparisonV2}.
One can see that in the PNJL model \cite{Ishii:2016dln}, in the scalar sector, the difference between the pole and screening masses below the critical temperature is much smaller — they both start to decrease smoothly around $130$ MeV, i.e., about $20$ MeV below $T_c$.
In the PNJL model \cite{Ishii:2016dln}, the pion masses start to increase near the phase transition, while in the quark-meson model \cite{Helmboldt:2014iya} the increase begins at $T\sim 100$ MeV.
Figure~\ref{MassesDiffComparison} shows the difference between the screening and pole masses. One can see that in the local PNJL model, this difference is always positive, while in the nonlocal model it is positive for the $\sigma$ meson but negative for the pion — i.e., the pole mass of the pion is larger than its screening mass. In the quark-meson model, this difference is also negative for the pion.
The ratio of the screening to pole masses is shown in Fig.~\ref{MassesRatioComparison}. In the nonlocal model, the mass difference is maximal near the phase transition, reaching up to about $30\%$. In the PNJL model, this occurs after the phase transition. In the quark-meson model, this ratio increases slowly in magnitude with temperature.    

\section{Conclusions}

In the previous paper, the confining extension of the nonlocal model was considered. It is based on a cut of the Laplace transform of the quark propagator with an additional dimensionful parameter and, to a large extent, leads to representing a part of the quark propagator as an entire function. In order to model the deconfinement phase transition at finite temperature, it was suggested to consider the quark propagator as a real quark pole with an additional entire function contribution. However, in this case, an inconsistency appears between the hadronic and quark matter phases. The temperature dependence of the four-quark coupling constants was introduced to compensate for this inconsistency.

Here, a different scheme for this compensation is proposed: the Laplace-transformed function is modified to reproduce the gap equation in the model without the confining extension. As an approximation, the original $\phi_3$ gap equation is used.
The mass spectrum of the $\pi$ and $\sigma$ mesons at finite temperature is studied, including both the screening and pole masses. 
The screening and pole masses of the pion and $\sigma$ meson are considered. The calculations of the pole masses are performed up to the deconfinement phase transition point, beyond which the mesons become unstable due to quark–antiquark dissociation, and no real-valued solution for the pole masses can be found. The pion pole mass is rather stable against temperature variations and only starts to increase in the vicinity of the phase transition. In contrast, the $\sigma$ meson pole mass begins to decrease already around $100$ MeV.

In the future, we plan to extend our model to finite chemical potential and include subleading $1/N_c$ terms. The $1/N_c$ corrections will introduce hadronic dressing with corresponding widths for unstable hadrons and, possibly, the dressing of quarks in quark matter.

This work is supported by supported by 
the CAS President’s International Fellowship Initiative (PIFI) No. 2026PVA0045 and No. 2023VMA0015; the project of the Ministry of Education and Science of the Russian Federation within the framework of the project
"Development  of  analytical  and  numerical methods of description in problems of mathematical physics, continuum mechanics, quantum field theory and nuclear physics" (no. of state registration: 126021217175-3); CAS Project for Young Scientists in Basic Research YSBR-088; the National Natural Science Foundation of China under Grants No. 12375117; and the Youth Innovation Promotion Association of Chinese Academy of Sciences (Grant No. Y2021414).

The authors thank the I.V. Anikin, O.V.Teryaev for fruitful comments.  %
\appendix
\section {Modification of quark propagator and pion vertex}

\begin{figure}[t]
    \centering
        \includegraphics[width=0.45\textwidth]{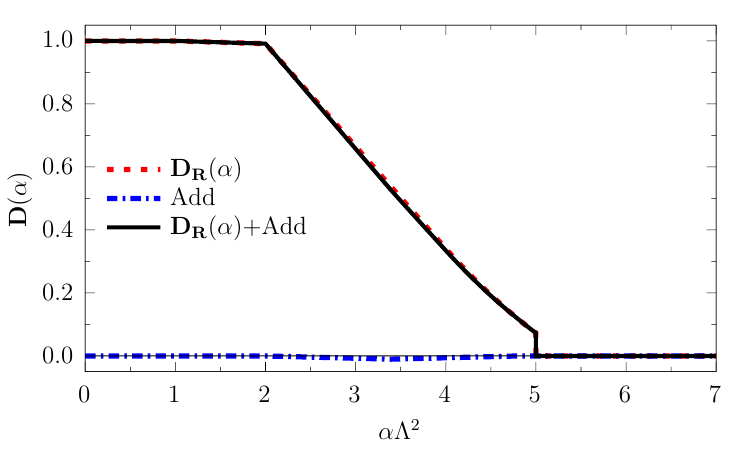}
\caption{Alpha-space function $\mathbf{D}(\alpha)$ obtained via the inverse Laplace transform of the scalar propagator in vacuum ($m_d=424$ MeV): the red dashed line represents $\mathbf{D}_R(\alpha)$ with $\Lambda_c^2 = \Lambda^2/5$, the blue dash-dotted line shows the additional contribution, and the solid line corresponds to the total expression.%
}
\label{LapImagVac}
\end{figure}

\begin{figure}[t]
    \centering
        \includegraphics[width=0.45\textwidth]{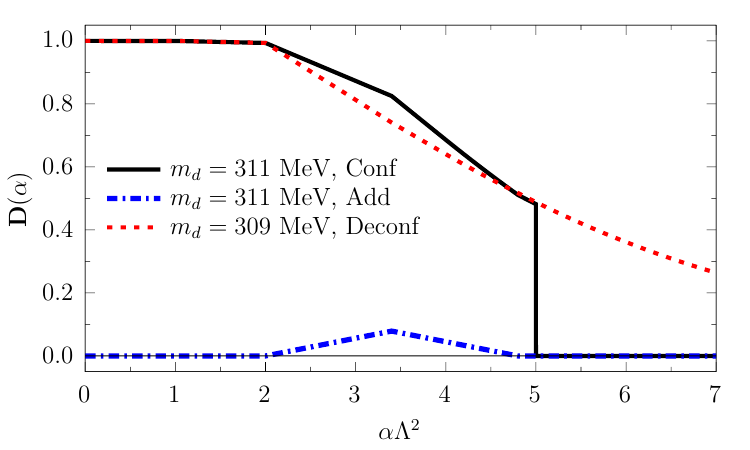}
\caption{Alpha-space function $\mathbf{D}(\alpha)$ obtained via the inverse Laplace transform of the scalar propagator at finite temperature near the phase transition: the solid line represents the total expression in the confinement phase ($m_d = 311$ MeV), with the blue dash-dotted line showing the additional contribution; the red dashed line corresponds to the deconfinement phase ($m_d = 309$ MeV).%
}
\label{LapImagPhaseTrans}
\end{figure}

The general idea of the modification is related to the transformation of the inverse Laplace transform of the scalar propagator \eqref{InvDRCon}. The function of the Laplace parameter $\alpha$ has the following properties: it behaves as $\exp(-m_c^2 \alpha)$ at small $\alpha$ and has a cut 
at $1/\Lambda_c^2$. In order to compensate for the contribution to the gap equation from the part that is cut off, one can introduce an additional function $D_R^{\text{Add}}(\alpha)$, which should lie between $\alpha_{\text{min}}$ and $\alpha_{\text{max}}$.
These particular values are restricted based on the requirement not to spoil the asymptotic freedom and confinement properties. The first condition suggests keeping the usual scalar propagator with the current quark mass at large space-like momenta. Since the first correction to it appears at $1/\Lambda^2$, the value of $\alpha_{\text{min}}$ should be at least larger than that scale. The second condition suggests that $\alpha_{\text{max}}$ be smaller than the cut value $1/\Lambda_c^2$. As the simplest function, a triangular pulse function is considered: it increases linearly from zero to a certain value $H_{\text{Add}}$ on the interval $[\alpha_{\text{min}}, \alpha_{\text{cen}}]$, then decreases linearly from this value to zero on $[\alpha_{\text{cen}}, \alpha_{\text{max}}]$, and vanishes elsewhere. The pulse is taken to be symmetric, i.e., $\alpha_{\text{cen}}=(\alpha_{\text{max}}+\alpha_{\text{min}})/2$. Specifically, the values $\alpha_{\text{min}}=2/\Lambda^2$ and $\alpha_{\text{max}}=4.8/\Lambda^2$ are chosen. The height $H_{\text{Add}}$ is fitted for a given $m_d$ to reproduce the value of the gap equation in vacuum or in medium without the confinement prescription. Based on general considerations, and in order for this triangular pulse to represent only a small correction, the condition $H_{\text{Add}}\ll 1$ should be satisfied. To illustrate the result, the inverse Laplace transforms $D_R(\alpha)$ and $D_R^{\text{Add}}(\alpha)$ are shown in Fig.~\ref{LapImagVac} for $m_d=424$ MeV (fitted in vacuum) and in Fig.~\ref{LapImagPhaseTrans} for $m_d=311$ and $309$ MeV (fitted at finite temperature). For $m_d=424$ MeV, the contribution $D_R^{\text{Add}}(\alpha)$ is hardly visible. The value $m_d=311$ MeV corresponds to the hadronic phase just below the phase transition. In this case, although $H_{\text{Add}}$ is maximal, $D_R^{\text{Add}}(\alpha)$ can still be regarded as a small perturbation. In the quark matter phase, with $m_d=309$ MeV, the leading asymptotic behavior is not cut off, so the total correction is very small.  

\begin{figure}[t]
    \centering
        \includegraphics[width=0.45\textwidth]{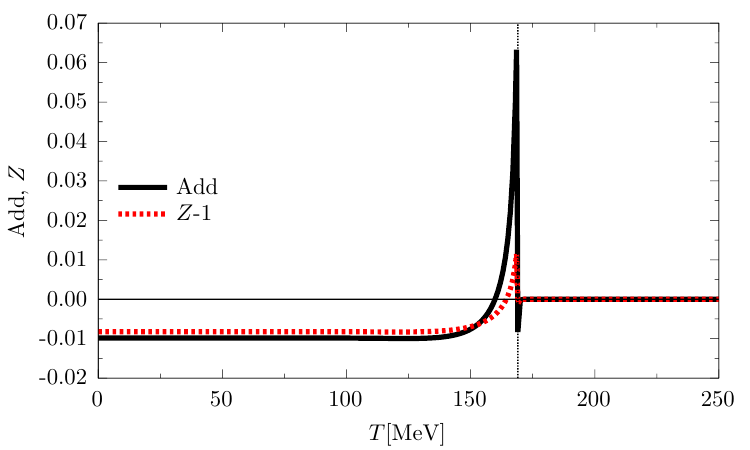}
\caption{Temperature dependence of the triangular pulse height $H_{\text{Add}}$ of the additional contribution $D_R^{\text{Add}}(\alpha)$ to the quark propagator and of the pion renormalization factor $Z$.}
\label{CorrectionDelZp}
\end{figure}

For a proper description of the pion properties, one needs to modify each vertex of the quark interaction by multiplying it by a $Z$ factor. With a constant $Z$, the screening mass of the pion would exhibit a discontinuity at the phase transition point, which seems unphysical even if it is small. Therefore, one can consider the case where $Z$ has an effective $T$ dependence and determine this dependence by fitting the pion screening mass to the result of the model with confinement. In Fig.~\ref{CorrectionDelZp}, the finite-temperature behavior of $H_{\text{Add}}$ and $Z-1$ is shown. One can see that both quantities exhibit a similar dependence: they are much smaller than $1$, meaning they represent only a correction; they retain their vacuum values up to $\sim130$ MeV, become maximal just near the phase transition, change sign at the phase transition point, and vanish after it.

\end{document}